\documentclass{iopart}
\usepackage{latexsym,hyperref,color,amssymb,graphicx}
\newcommand{\beq}{\begin{equation}}
\newcommand{\eeq}{\end{equation}}
\newcommand{\bea}{\begin{eqnarray}}
\newcommand{\eea}{\end{eqnarray}}
\newcommand{\bean}{\begin{eqnarray*}}
\newcommand{\eean}{\end{eqnarray*}}
\newcommand{\bei}{\begin{itemize}}
\newcommand{\eei}{\end{itemize}}
\newcommand{\ben}{\begin{enumeration}}
\newcommand{\een}{\end{enumeration}}
\newcommand{\nn}{\nonumber}

\definecolor{darkorange}{rgb}{.6,.2,.0}

\definecolor{darkgreen}{rgb}{0.0,0.7,0.0}

\begin{document}

\title{Alternative splicing and genomic stability}

\author{Kevin Cahill}
\ead{cahill@unm.edu}
\address{Department of Physics and Astronomy, 
University of New Mexico, Albuquerque, NM 87131, USA}
\address{Center for Molecular Modeling, 
Center for Information Technology, 
National Institutes of Health, 
Bethesda, Maryland 20892-5624, USA}

\date{\today}

\begin{abstract}
Alternative splicing
allows an organism 
to make different proteins
in different cells at
different times, all
from the same gene.
In a cell that uses alternative splicing, 
the total length of all the exons 
is much shorter than in a cell that encodes
the same set of proteins without 
alternative splicing.
This economical use of exons
makes genes more stable 
during reproduction and development
because a genome with a shorter exon length 
is more resistant to harmful mutations.
Genomic stability may be the reason why
higher vertebrates splice alternatively.
For a broad class
of alternatively spliced genes,
a formula is given for the increase
in their stability.
\par
\vspace{0.2in}
\par\noindent
Keywords: splicing, alternative splicing, exon, genomic stability.
\end{abstract}

\pacs{87.10.+e, 87.14.Gg, 87.17.Ee, 87.23.Kg}

\maketitle
What is alternative splicing?
A procaryote (no nucleus)
transcribes one or more genes into mRNA 
and immediately translates the mRNA into protein.
But a eucaryote 
first transcribes a single gene into
pre-mRNA, and then, using 
spliceosomes, turns the pre-mRNA 
into mRNA by
splicing out most
or all of its introns and often many
of its exons.
The eucaryote then exports the mRNA out
of its nucleus into its cytosol,
where its ribosomes translate 
the mRNA into protein.
A eucaryote often can make different proteins
from the same pre-mRNA transcript 
by splicing it in different ways.
This trick is called \textit{alternative splicing}\@.
\par
Why do higher vertebrates splice alternatively?
Alternative splicing
allows an organism 
to make different proteins
in different cells at
different times, all
from the same gene,
by poorly understood regulatory devices~\cite{Alberts2002}\@.
But this diversity of proteins 
could also be produced by several
different genes controlled by
promoters and enhancers --- in fact,
that is how biologists thought genes
worked until they discovered alternative splicing.
The advantage of alternative splicing 
is that its economical use of exons
makes genes more stable 
during reproduction and development.
\par
This communication gives 
a formula and a rule of thumb for the increase
in the stability of a broad class of 
alternatively spliced genes.
The \textit{DSCAM} gene of the fruit fly
and the \textit{cSlo} gene of the chicken 
provide examples that 
illustrate the formula and the rule.
A Monte Carlo simulation, displayed
in Figure 1, suggests how
alternative splicing may help
dividing human cells
avoid excessive mutations.
\par
How does alternative splicing make genes more stable?
Consider, for instance, 
a gene that has a long
exon of 1000 base pairs (b)
and two short ones, each  
100 b long.  The total 
length of its exons is 1200 b.
Alternative splicing
allows the cell to make two 
different RNAs, each of 1100 b.
Without alternative splicing, 
the cell would need two genes,
each 1100 b long,
for a total exon length of 2200 b.
Thus in this example,
alternative splicing reduces the length
of the exons in the DNA by 45\%\@. 
This reduction in the length
of exonic DNA implies
a reduction of 45\% in the error rate
during the replication of this gene.
In effect, the gene is nearly twice
as stable due to alternative splicing.
Since an error in the replication
of critical exonic DNA
is potentially lethal,
this extra genomic stability
is biologically significant and
is one of the reasons
why higher eucaryotes use 
alternative splicing.
Computer scientists will recognize
alternative splicing as akin to
file compression~\cite{Ford2001}.
\par
More generally,
let us consider a gene that has \(M\) 
groups of mutually exclusive exons
in addition to the constitutively spliced exons
(the exons that are always kept in the mRNA)\@.
For each group \(i\; (i=1, \dots M)\),
let  \(N_i\) denote the number
of mutually exclusive exons in the group,
including the null exon of length zero
if the spliceosome may splice out
all the exons of the group.
Assume that the spliceosome
always selects at most one exon and no introns 
from each of the \(M\) groups of \(N_i\) exons
with no shuffling.
Assume that the organism expresses all 
\beq
\mathcal{N} = \prod_{j=1}^M N_j
\label{Nprot}
\eeq
possible proteins at some time 
in some cell. 
\par
Let us use \(L_\mathrm{c}\) for the total length 
in nucleotides of the constitutively spliced exons. 
Without alternative splicing,
these \(L_\mathrm{c}\) nucleotides would be repeated
in each of the \(\mathcal{N}\) proteins for
a total length of \(\mathcal{N} \, L_\mathrm{c}\)\@.
\par 
If \(L_{ik}\) is the length of exon \(k\)
of group \(i\),
then without alternative splicing,
each of the \(N_i\) exons of length \(L_{ik}\) 
would be repeated \(\mathcal{N}/N_i\) times.
So the total length devoted to group \(i\)
without alternative splicing would be
\beq
\frac{\mathcal{N}}{N_i} \,
\sum_{k=1}^{N_i} \, L_{ik}
=
\left( \prod_{j = 1 \atop j\ne i}^M N_j \right)
\, \sum_{k=1}^{N_i} \, L_{ik} .
\label{NLi}
\eeq
\par
Thus the number of nucleotides that would
be needed to encode all \(\mathcal{N}\)
proteins and that would
have to be copied correctly each time
a cell divides is
\beq
N_{\mathrm{nas}} = \mathcal{N}\, \left(
L_\mathrm{c} + \sum_{i = 1}^M \, \frac{1}{N_i} \,
\sum_{k=1}^{N_i} \, L_{ik} \right)
\label{Nnas}
\eeq
 without alternative splicing.
\par
But with alternative splicing,
the number of needed nucleotides is only
the length of all the exons,
\beq
N_{\mathrm{as}} = L_\mathrm{c} + \sum_{i=1}^M \, \sum_{k=1}^{N_i} \, L_{ik} .
\label{Nas}
\eeq
\par
Since the error rate in the replication of DNA
is \(10^{-9}\) per base pair~\cite{Alberts2002},
the probability of an exonic error in the gene during 
replication is \(N_{\mathrm{nas}} \, \times \, 10^{-9}\)
without alternative splicing,
but only \(N_{\mathrm{as}} \, \times \, 10^{-9}\) with
alternative splicing.
So if we ignore the critical control
sequences in the introns, then
the ratio 
\beq
I = \frac{N_{\mathrm{nas}}}{N_{\mathrm{as}}}
\label{Idef}
\eeq 
is the increase in the stability of the gene
due to alternative splicing.
The intron control sequences probably
boost \(I\) slightly.
\par
The \textit{DSCAM} gene of \textit{Drosophila}
provides a striking example of alternative splicing.
This gene encodes receptors that 
guide the growth of the axon of Bolwig's nerve 
in the fly embryo~\cite{Schmucker2000}\@.
It has \(M = 4\) groups of
\(N_1 = 12\),
\(N_2 = 48\), \(N_3 = 33\), 
and \(N_4 = 2\) exons~\cite{Schmucker2000,Black2000}\@. 
The exons in each group are mutually exclusive,
and the total number of possible proteins is
\(\mathcal{N} = 12\times48\times33\times2 = 38,016\)\@.
The \textit{DSCAM} gene, including introns, is 61.2 kb long,
and its mRNA, after transcription and splicing, 
contains 24 exons and is 7.8 kb 
long~\cite{Schmucker2000,Black2000}\@.
\par
The ratio \(N_{\mathrm{nas}}/N_{\mathrm{as}}\) depends
explicitly upon the lengths \(L_\mathrm{c}\) 
and \(L_{ik}\)\@.
Since most internal exons are between 50 
and 300 nucleotides in length~\cite{Smith2000},
let us simplify the bookkeeping by
setting \(L_{ik} = 200\) b.
The spliced \textit{DSCAM} mRNA is 7.8 kb long and 
contains 4 alternatively spliced exons and 20
constitutively spliced exons.
So the set of constitutively spliced exons 
is of length
\beq
L_\mathrm{c} = 7800 \, - \, 4\times200 = 7000
\label{Lr}
\eeq
or \(L_\mathrm{c} = 7\) kb.
Thus by Eq.(\ref{Nas}), 
the exonic length required with alternative splicing is
\bea
N_{\mathrm{as}} & = & L_\mathrm{c} + 200 \, \sum_i^4 N_i\nn\\
& = & 7000 + 200 \, (12 + 48 + 33 + 2) = 26000
\label{Nasd}
\eea
or \(N_\mathrm{as} = 26\) kb.
But by Eq.(\ref{Nnas}), 
the exonic length required 
without alternative splicing is
\bea
N_{\mathrm{nas}} &
= & \mathcal{N} \, ( L_\mathrm{c} \, + \, 200 \, M ) \nn\\
& = & 38016 \times 7800 = 296524800
\label{Nnasd}
\eea
or \(N_{\mathrm{nas}} = 297\) Mb, which,
incidentally, is nearly
twice the length of the entire 
\textit{Drosophila} genome
and about six times the length
of all the exons in the human genome.
With these assumptions, 
the chance of a crucial error 
in the \textit{DSCAM} gene during
replication is 
\(0.30\)
without alternative splicing, but only
\(2.6 \, \times \, 10^{-5}\)
with alternative splicing.
The ratio \(I\)
\beq
I = \frac{N_\mathrm{nas}}{N_\mathrm{as}} 
= 11400 = 1.1 \times 10^4
\label{Id}
\eeq
is the increase
in genetic stability due to alternative splicing.
\par
Fruit flies without alternative splicing 
would accumulate about 10,000 exonic \textit{DSCAM} errors
in 30,000 generations (2,500 years),
and each fly would have its own set of 10,000 errors. 
Over this period, the \textit{DSCAM} gene of
the fly population gradually would become 
uniformly dysfunctional with relatively small
differences in fitness among individual flies.
On the other hand,
flies with alternative splicing 
would accumulate less than one exonic \textit{DSCAM} error
in 30,000 generations.
Moreover, the probability that the one error
would occur in the \(L_\mathrm{c}\) exons 
that are constitutively expressed 
would be \(L_\mathrm{c}/N_\mathrm{as} = 7/26 =0.27\),
and that unlucky fly would be distinctly unfit.
Thus alternative splicing not only 
avoids exonic errors; it also helps natural selection 
weed out unfit individuals. 
Alternative splicing and natural selection 
cooperate to preserve the integrity of the genome.
\begin{figure}
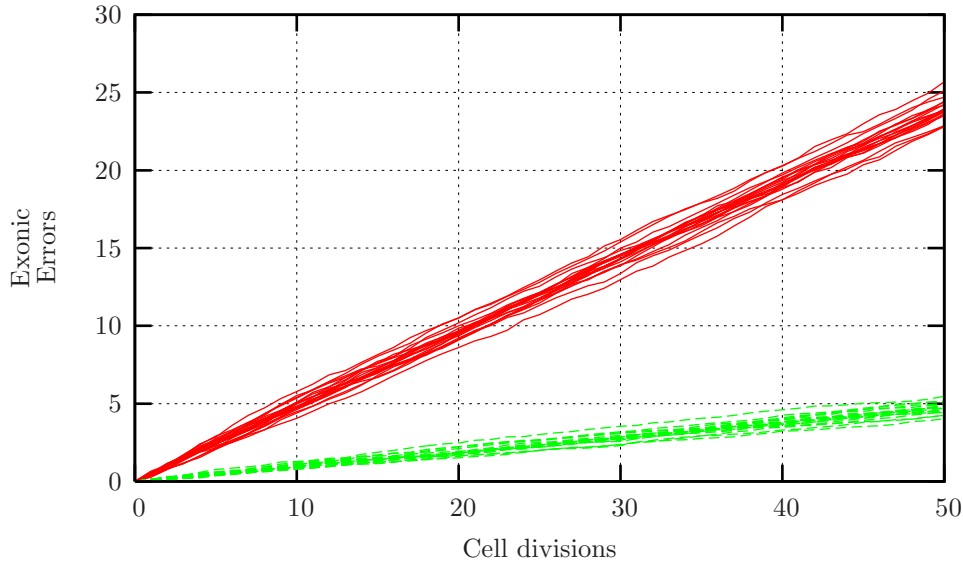

\centering
\input DEFECTS
\caption{After 46 cell divisions,
the number of defects in 
the exons of a human diploid cell
increases to \(4.4 \pm 0.07\)
with alternative splicing
(lower curves, green) and to
\(22.0 \pm 0.14\)
without alternative splicing
(upper curves, red).}
\label{defects}
\end{figure}
\par
In most genes, the increase in genomic
stability due to alternative splicing
might be more like 5 or 10 than \(10^4\),
but even a 500\% increase in genetic stability 
during reproduction and development
is worth the
trouble of alternative splicing.
For if without alternative splicing
the average gene were 5 times longer,
then 7.5\% rather then 1.5\% of the genomes
of higher vertebrates
would consist of exons.  The DNA of a 
human diploid cell
has 6.4 billion base pairs.
The error rate of \(10^{-9}\) per base pair
implies that on average there will be 6.4
errors per cell division.
With alternative splicing, only 1.5\%
of these errors occur in exons 
and are potentially deleterious,
so the probability of a daughter cell with
perfect exons 
is approximately \(P_{\mathrm{as}} = 1 - 6.4 \times 0.015
= 0.904\)\@.  
A more accurate estimate is
\beq
P_{\mathrm{as}} = (1 - 10^{-9})^{0.015\times6.4\times10^9}
\approx e^{-0.096} = 0.908.
\label{Pas}
\eeq
Without alternative splicing,
7.5\% of the errors would occur in exons,
and so the probability of a daughter cell with
perfect exons would be roughly
\(P_{\mathrm{nas}} = 1 - 6.4 \times 0.075
= 0.52\)\@.
A more accurate estimate is
\beq
P_{\mathrm{nas}} = (1 - 10^{-9})^{0.075\times6.4\times10^9}
\approx e^{-0.48} = 0.619.
\label{Pnas}
\eeq 
The adult human arises from about
46 cell divisions, so the probability
that any given adult cell has perfect exons
is \((P_{\mathrm{as}})^{46} = 0.012\)
with alternative splicing, but only
 \((P_{\mathrm{nas}})^{46} = 2.6 \times 10^{-10}\)
without alternative splicing.
\par
To estimate the implications of alternative splicing
for human evolution and development,
I again assumed that the human genome
without alternative splicing would
have five times more exonic base pairs.
I let two sets of 1000 cells divide 50 times
\textit{in silico}\@.
The set of cells that used alternative splicing
had \(0.015\times6.4\times10^9 = 9.6\times10^7\) 
exonic base pairs; the set that did not use
alternative splicing had 5 times as many or
\(4.8\times10^8\) exonic base pairs.
I divided the 1000 cells into 20 groups
of 50 cells each and
plotted in the figure the average number of exonic errors
per cell for each of the 20 groups
with and without alternative splicing.
As shown in the figure,
the average number of defective exonic base pairs
per daughter cell after 46 cell divisions is
\(4.43 \pm 0.07\) with alternative splicing (lower, green lines)
but \(22.0 \pm 0.14\) without alternative splicing 
(upper, red lines).
Since with alternative splicing,
cells free of exonic error
produce daughter cells that also
are free of exonic error at a rate of 91\%,
apoptosis followed by division
of adjacent cells can correct the 
1 or 2 of the 4 exonic errors that 
are troublesome.
But because without alternative splicing,
cells free of exonic error
produce daughter cells 
free of exonic error at a rate of
only 62\%, it is hard to see how
apoptosis could cope with 22 exonic errors
per adult cell.
\par
We may derive a rule of thumb
for the increase in genetic stability
by noting that \(\langle L_\mathrm{s} \rangle\)
defined by
\beq
\langle L_\mathrm{s} \rangle = 
\sum_{i = 1}^M \, \frac{1}{N_i} \,
\sum_{k=1}^{N_i} \, L_{ik}
\label{Ls}
\eeq
is an effective average length of the alternative
exons that are spliced into the mRNA and that
\(\langle N \rangle\) defined by 
\beq
\langle N \rangle \, \langle L_\mathrm{s} \rangle = 
\sum_{i=1}^M \, \sum_{k=1}^{N_i} \, L_{ik}
\label{barN}
\eeq
is a kind of
average of the numbers \(N_i\)
of alternative exons in the \(M\) groups.
Let us further use \(r\) for the ratio 
of the average length
\( \langle L_\mathrm{s} \rangle \) 
of the selected exons to
the length \(L_\mathrm{c}\)
of the constitutively spliced exons
\beq
r = \frac{\langle L_\mathrm{s} \rangle}{L_\mathrm{c}}.
\label{r}
\eeq
Then with these definitions,
the increase \(I\) in genetic stability is
\beq
I = \frac{N_{\mathrm{nas}}}{N_{\mathrm{as}}}
= \mathcal{N} \, \frac{1 + r}{1 + \langle N \rangle \, r}.
\label{I}
\eeq
\par
The fraction that multiplies
the total number \( \mathcal{N} \)
of possible proteins is less than unity.
But it is generally not tiny because 
the ratio \(r\) usually is small
and because \( \langle N \rangle \)
usually is less then 30.
In the case of \textit{Drosophila} \textit{DSCAM} and
with the assumptions \(L_\mathrm{c} = 7.0\) kb 
and \(L_{ik} = 200\) b, 
the four selected exons are of length
\(\langle L_\mathrm{s} \rangle = 800\) b;
the ratio \(r\) is
\(r = 800/7000 = 0.114\); and the effective
average number \(\langle N \rangle\) 
of exons per group is
\(\langle N \rangle = 
95 \times 200/\langle L_\mathrm{s} \rangle = 95/4
= 23.7\).
The fraction \((1+r)/(1 + \langle N \rangle \, r) = 3/10\),
and the increase in genetic stability is
\(I = 0.3 \, \mathcal{N} = 11,400\)\@.
\par
Hearing in chickens provides another example
of the contribution of alternative splicing
to genetic stability.
The \textit{cSlo} gene of the chicken cochlea
encodes the membrane proteins that form 
the Ca\(^{2+}\)-activated K\(^+\) 
channels that determine
the resonant frequency of
each hair cell in the basilar papilla.
Alternative splicing provides some
\(\mathcal{N} = 576\) variants of the
mRNA for each of the four
components of this tetramer membrane 
protein~\cite{Hudspeth1997,Oberholtzer1997,Black1998},
resulting in a huge number
possible resonant frequencies.
In \textit{cSlo}, the ratio \(r = 0.1\), 
and the mean number \(\langle N \rangle\) of exons
in each of the eight groups is about 
2.6~\cite{Hudspeth1997}\@.
So by the rule of thumb (\ref{I}), alternative splicing
increases the genetic stability of \textit{cSlo}
by a factor of about
\beq
I = 576 \, \frac{1.1}{1.26} = 503.
\label{Icslo}
\eeq
The tetrameric structure
of the functional membrane
protein effectively boosts \(I\) by another
factor. 
\par
Another example of alternative splicing's exonic economy 
is provided by the mammalian immune system,
which uses site-specific genetic 
recombination in developing B cells~\cite{Alberts2002}\@.
\par
We have seen that the exonic economy
of alternative splicing increases the
stability of the genome.
As genomics and proteomics advance,
the protein-to-gene
ratios of the higher vertebrates will teach us
how much alternative splicing actually
contributes to the stability of their genomes.

\ack Thanks to Mark Fleharty,
Michael Malik, Valerie Parsegian,
Stephanie Ruby, and James Thomas 
for helpful conversations.
Peter J.~Steinbach kindly hosted
the author during his sabbatical at NIH\@.

\section*{References}
\bibliography{bio,cs}
\end{document}